\documentclass[a4paper,11pt]{article}
\usepackage{pos}

\newcommand{\bea}{\begin{eqnarray}}
\newcommand{\eea}{\end{eqnarray}}
\newcommand{\bean}{\begin{eqnarray*}}
\newcommand{\ean}{\end{eqnarray*}}

\newcommand{\be}{\begin{equation}}
\newcommand{\ee}{\end{equation}}
\newcommand{\eq}[1]{Eq.~(\ref{#1})}
\newcommand{\fig}[1]{Fig.~\ref{#1}}
\newcommand{\tab}[1]{Table~\ref{#1}}

\newcommand{\ev}[1]{\left\langle #1 \right\rangle}
\newcommand{\Op}{\mathcal{O}}

\title{Mixing of heavy and light quarks in charmonium and light mesons}
 \ShortTitle{Flavor mixing in charmonium and light mesons}

\author*[a]{Francesco Knechtli}
\author[b]{Jacob Finkenrath}
\author[a]{Roman H{\"o}llwieser}
\author[a]{Tomasz Korzec}
\author[c]{Michael Peardon}
\author[a]{Juan Andrés Urrea-Niño}

\affiliation[a]{Department of Physics, University of Wuppertal,\\
  Gau{\ss}stra{\ss}e 20, 42119 Wuppertal, Germany}
\affiliation[b]{CERN,\\
  Esplanade des Particules 1, 1211 Geneva 23, Switzerland}
\affiliation[c]{School of Mathematics, Trinity College Dublin,\\
Dublin 2, Ireland}

\emailAdd{knechtli@uni-wuppertal.de}

\abstract{We study the system of light mesons, charmonium and glueballs in the flavor singlet scalar channel where they can mix. We use lattice QCD simulations with an almost physical charm quark and three degenerate light quarks for two values of the pion mass ($m_{\pi} \approx 420, 800$ MeV). Thanks to a variational basis which includes mesonic operators with profiles in distillation space, Wilson loops and two-pion operators we detect and show results of their mixing.}

\FullConference{The XVIth Quark Confinement and the Hadron Spectrum Conference (QCHSC24)\\
 19-24 August, 2024\\
 Cairns Convention Centre, Cairns, Queensland, Australia\\}

\begin{document}
\maketitle

\section{Motivation}

Confinement predicts the existence of states made of gluons alone called
glueballs which still await experimental confirmation. There is renewed interest both
from the experimental~\cite{BESIII:2023wfi} and the theoretical side~\cite{Morningstar:2024vjk}. Yet another ``puzzle'', more exotic states than ``just'' mesons and baryons were discovered called the X, Y and Zs. Despite experimental discoveries two decades ago there is still no understanding of their internal mechanics. For example the $J^{PC}=1^{++}$ channel contains $\chi_{c1}(3872)$ (formely called $X(3872)$), one of the longest standing candidates for a state which is not simply a charmonium state made of a charm quark-anti-quark pair. The nature of this state is still being investigated~\cite{Ji:2025hjw}.

Glueballs are expected at energies close to charmonium. In QCD they are resonances with many decay channels e.g. into pions. Several XYZ states have been discovered with charmonium content. This motivates to study the mixing of light and charm quarks. Lattice QCD provides an ab-initio tool to study glueballs and exotics but it is challenging. For example gluonic observables or observables with quark-line disconnected contributions suffer from a signal-to-noise problem. Moreover glueballs and the XYZ states are resonances which decay by the strong interactions. Taking into account that these particles are unstable is not straightforward in a lattice calculation.

\section{Short introduction to lattice QCD}

Lattice QCD is very good at computing Euclidean correlation functions
\be
  \ev{ \Op(t)\Op^\dagger(0) } \,, \label{e:vev}
\ee
where the ``operator'' $\Op$ is a (temporally) local combination of fields.
The integral expression \eq{e:vev} is related to an underlying QFT with Hamiltonian $\hat H$ and complete set of energy eigenstates $|n\rangle$, $\hat H |n\rangle = E_n |n\rangle$, $n=0,1,2,\ldots$. The energy eigenstates can be chosen to be orthonormal
\be
 \langle n|m\rangle = \delta_{n,m}\,,\quad \sum_n |n\rangle \langle n| = 1 \,.
\ee
A consequence of this relation is the ``spectral decomposition''
\be
 \ev{ \Op(t)\Op^\dagger(0) } = \sum_n |c_n|^2 e^{-E_n t}
\ee
with matrix elements between the eigenstate and the vacuum state $|\Omega\rangle$ called \emph{overlaps}
\be
 c_n = \langle n| \Op^\dagger |\Omega\rangle \,.
\ee
The energy eigenvalues $E_n$, including excited states $n>0$ can be efficiently extracted by the GEVP (Generalized EigenValue Problem) method \cite{Luscher:1990ck}. It starts by considering several operators for the same symmetry channel $\Op_1,\ldots, \Op_{N_{\text{op}}}$ and computing a correlation matrix with elements
\be
 C_{ij}(t) = \ev{ \Op_i(t) \Op^\dagger_j(0) } \label{e:vevmatrix}
\ee
The generalized eigenvalues $\lambda_n(t,t_{\rm ref})$ of
\be
 C(t)\; v_n(t,t_{\rm ref}) = \lambda_n(t,t_{\rm ref})\; C(t_{\rm ref})\; v_n(t,t_{\rm ref}) \label{e:gevp}
\ee
behave like
\be
 \lambda_n(t,t_{\rm ref}) = {\rm e}^{-(t-t_{\rm ref})E_n} \times \left(1 + {\rm e}^{-(t-t_{\rm ref})\Delta_n}\right) \,,
\ee
where the value of $\Delta_n$ depends on conditions on $t$ and $t_{\rm ref}$ as discussed in \cite{Blossier:2009kd}. For the GEVP method to work we need a variational basis with operators that have good overlaps \cite{Dudek:2010wm}
\be
 \langle n| \Op_i^\dagger | \Omega \rangle \propto \left[ C(t_0) v_n(t,t_0)  \right]_i \label{e:overlap}
\ee
with all states $|n\rangle$.

The operators $\Op_i$ in \eq{e:vevmatrix} must have the same symmetries. Besides the total angular momentum $J$, the parity $P$ and the charge conjugation $C$ there is the light flavor symmetry. Nature has no exact flavor symmetry, but an approximate $SU(2)$, or even $SU(3)$. In $N_{\text f}=3+1$ QCD, which we consider here,
\be
 \begin{pmatrix} u\\ d\\ s\end{pmatrix} \to V\begin{pmatrix} u\\ d\\ s\end{pmatrix},
 \quad (\bar u, \bar d, \bar s) \to (\bar u, \bar d, \bar s) V^\dagger,\qquad V\in SU(3)
\ee
is a symmetry. This means that energy eigenstates are labeled by $|D,Y,I,I_3 \rangle$ \cite{McNamee:1964xq}. In the $SU(3)$ flavor limit there is for example an octet $D=8$ of degenerate pseudoscalar mesons ($\pi^0$, $\pi^\pm$, $\eta$, $K^0$, $\overline{K^0}$, $K^\pm$) and a singlet $D=1$ ($\eta'$).  

A scattering process for example of two incoming and outgoing particles happens in real time. There are very short lived hadrons which are resonances forming during the scattering process and decaying via the strong interaction. Examples are the scalar singlet mesons $\sigma=f_0(500)$ and $f_0(980)$. The latter can be observed in proton-proton collisions and decays into a pair of pions $\pi^+\pi^-$ \cite{Lorenzo:2019qax}. Lattice QCD is formulated in Euclidean (imaginary) time though. While this is perfect for spectroscopy there is no direct access to scattering. L{\"u}scher's idea to circumvent this problem is to compute the discrete two-particle spectrum in a finite volume and solve an equation to find the phase shifts and infer the resonance parameters in infinite volume \cite{Luscher:1990ux}.

\begin{figure}[t]
  \centering
  \includegraphics[width=0.25\textwidth]{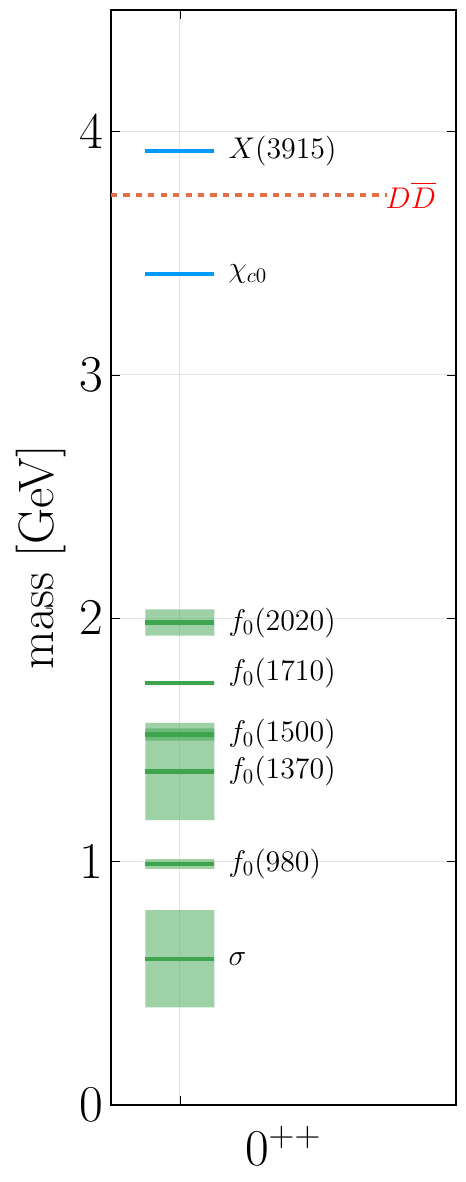}
  \hspace{0.1\textwidth}
  \includegraphics[width=0.25\textwidth]{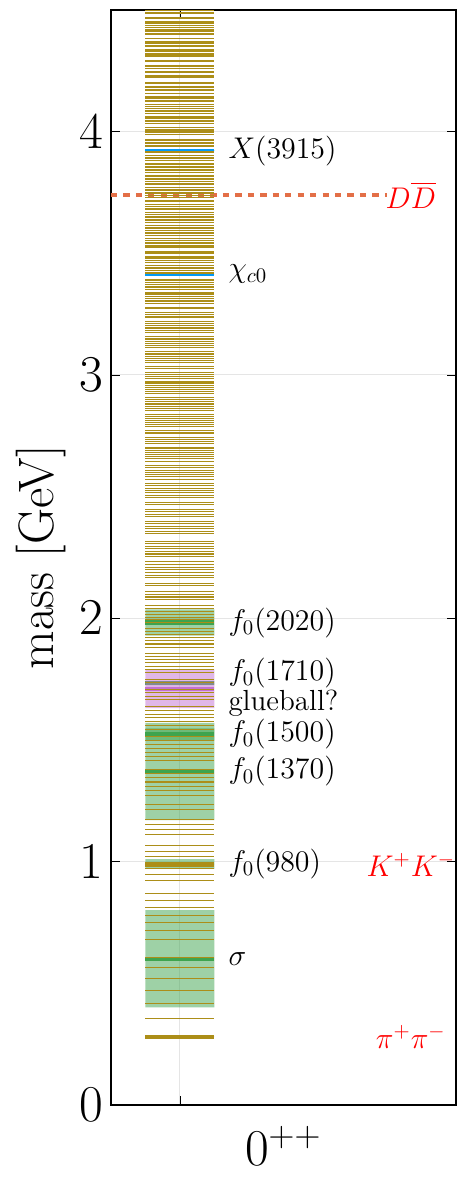}
  \hspace{0.1\textwidth}  
  \includegraphics[width=0.25\textwidth]{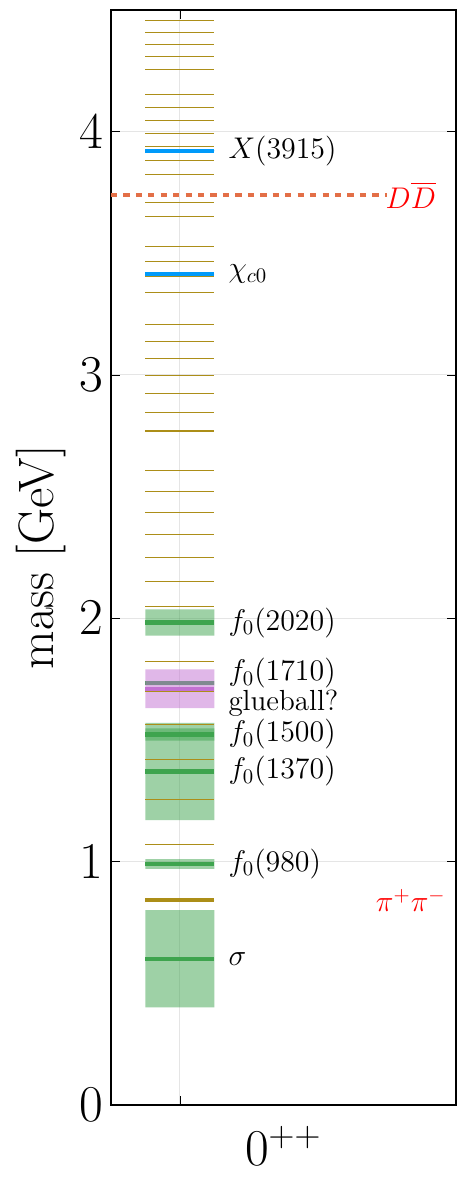}
  \caption{The spectrum of scalar mesons in nature (left, only one particle states are shown), in finite volume with physical pions (center) and heavier pions (right). Courtesy of T. Korzec.
    \label{f:scalarspectrum}
  }
\end{figure}
In this work we focus on the spectrum of scalar ($J^{PC}=0^{++}$) mesons which are singlets ($D=1$) under the flavor symmetry. We are interested in the mixing between charmonium and light meson states and we want to look for the existence of a glueball state. On the left plot of \fig{f:scalarspectrum} we show the experimental spectrum of single particle states from the Particle Data Group \cite{ParticleDataGroup:2022pth}. Lattice calculations in pure gauge theory yields a glueball state at $1710(50)(80)\,$MeV \cite{Chen:2005mg}. In a theory with dynamical fermions the glueball is unstable and decays into pions (up to 10) . If we include two-meson states $\pi\pi$, $KK$, $\ldots$ and also with momentum $\pi(\vec p)\pi(-\vec p)$ we get a continuum of states above the energy of a $\pi^+\pi^-$ state. Putting everything in finite volume helps. Momenta are quantized $\vec p=\vec n\ 2\pi/L$ (free case) and one gets a situation illustrated by the central plot of \fig{f:scalarspectrum}. The situation is further simplified if we consider heavier pions. E.g. at the $SU(3)$ flavor symmetric point, where up, down and strange quarks are degenerate but the sum of their masses is like at the physical point, the pion mass is $m_\pi=420\,$MeV. The finite volume spectrum in this situation is shown in the right plot of \fig{f:scalarspectrum}. We will follow the strategy to simulate QCD with $N_{\text f}=3+1$ flavors of quarks with a light quark mass corresponding to $m_\pi\approx800\,$MeV and then lower the light quark mass to $m_\pi\approx420\,$MeV. In this way we want to learn about the fate of the glueball state.

Our operator basis in the scalar flavor-singlet channel consists of
\begin{itemize}
\item scalar glueball\\
  we choose $\Op_{g}$ to be the sum of Laplacian eigenvalues \cite{Morningstar:2013bda}. This operator yields a slightly better (but comparable) signal for the correlator than a linear combination of Wilson loops of up to 35 different shapes and lengths, see Appendix A of \cite{Barca:2024fpc};
\item charmonium
  \be
  \Op_{c} = \bar c c
  \ee
 \item light scalar
   \be
   \Op_{l} = \frac{1}{\sqrt{3}}\left(\bar u u+ \bar d d + \bar s s \right)
   \ee
 \item two-pion operators with both particles at rest
   \bea
   \Op_{2\pi} = 
   -\frac{1}{\sqrt{8}}\left( \bar u \gamma_5 s \bar s \gamma_5 u + \bar d \gamma_5 s \bar s \gamma_5 d + \bar u \gamma_5 d \bar d \gamma_5 u + \bar d \gamma_5 u \bar u \gamma_5 d + \bar s \gamma_5 d \bar d \gamma_5 s 
   + \bar s \gamma_5 u \bar u \gamma_5 s \right) \nonumber \\
   -\frac{2}{3\sqrt{8}}\left(\bar u \gamma_5 u \bar u \gamma_5 u + \bar d \gamma_5 d \bar d \gamma_5 d + \bar s \gamma_5 s \bar s \gamma_5 s \right) \nonumber \\
   +\frac{1}{3\sqrt{8}}\left(\bar u \gamma_5 u \bar d \gamma_5 d + \bar d \gamma_5 d \bar u \gamma_5 u + \bar u \gamma_5 u \bar s \gamma_5 s + \bar d \gamma_5 d \bar s \gamma_5 s + \bar s \gamma_5 s \bar u \gamma_5 u + \bar s \gamma_5 s \bar d \gamma_5 d \right)
   \eea
\end{itemize}
The operators are projected to zero spatial momentum on a given timeslice $t$. The correlation matrix \eq{e:vevmatrix} is
\be
\begin{pmatrix}
\left \langle \Op_{l}(t) \bar{\Op}_{l}(0) \right \rangle & \left \langle \Op_{l}(t) \bar{\Op}_{c}(0) \right \rangle & \left \langle \Op_{l}(t) \bar{\Op}_{2\pi}(0) \right \rangle & \left \langle \Op_{l}(t) \bar{\Op}_{g}(0) \right \rangle\\
* & \left \langle \Op_{c}(t) \bar{\Op}_{c}(0) \right \rangle & \left \langle \Op_{c}(t) \bar{\Op}_{2\pi}(0) \right \rangle & \left \langle \Op_{c}(t) \bar{\Op}_{g}(0) \right \rangle \\
* & * & \left \langle \Op_{2\pi}(t) \bar{\Op}_{2\pi}(0) \right \rangle & \left \langle \Op_{2\pi}(t) \bar{\Op}_{g}(0) \right \rangle \\
* & * & * & \left \langle \Op_{g}(t) \bar{\Op}_{g}(0) \right \rangle
\end{pmatrix}\,,\label{e:corrmatrix}
\ee
where $*$ stands for the transpose element. We smear the quark fields using distillation \cite{HadronSpectrum:2009krc} and introduce profiles in distillation space as in \cite{Knechtli:2022bji} to optimize the overlaps onto the physical states.

\section{Mixing of flavors, glueballs and 2-pion states in the scalar channel}

The results presented here have been computed on gauge configurations of $N_{\rm f}=3+1$ QCD. They were generated by us with a Wilson fermion action with non-perturbatively determined mass-dependent clover improvement \cite{Fritzsch:2018kjg} and a tree level improved L\"uscher--Weisz gauge action \cite{Luscher:1984xn}. The parameters of the ensembles are listed in \tab{t:ensembles}. The ensemble at the SU(3) flavor symmetric point has been generated in \cite{Hollwieser:2020qri}. The ensemble with the heavier pion is newer and has been recently used in \cite{Urrea-Nino:2025ztj,Hollwieser:2025nvv,Struckmeier:2025ebs}.

\begin{table}
\begin{center}
\begin{tabular}{||c c||} 
 \hline
 A1 & A1h \\ [0.5ex] 
 \hline\hline
 $96\times 32^3$ & $96\times 32^3$ \\ 
 \hline
 $a \approx 0.054\ \text{fm}$  & $a \approx 0.069	\ \text{fm}$ \\
 \hline
 $m_{\pi} \approx 420\ \text{MeV}$  & $m_{\pi} \approx 800\ \text{MeV}$ \\
 \hline
 $N_v^{\rm light} = 100$ & $N_v^{\rm light} = 200$ \\
 \hline
 $N_v^{\rm charm} = 200$ & $N_v^{\rm charm} = 200$ \\ [1ex] 
 \hline
\end{tabular}
\end{center}
\caption{Parameters of the $N_{\rm f}=3+1$ QCD gauge ensembles used in this study. In order the rows show: the lattice size, the lattice spacing determined from the $h_c-\eta_c$ mass splitting, the pion mass, the number of distillation vectors used to smear the light and the charm quark field respectively. \label{t:ensembles}}
\end{table}

\begin{figure}[t]
  \centering
  \includegraphics[width=0.7\textwidth]{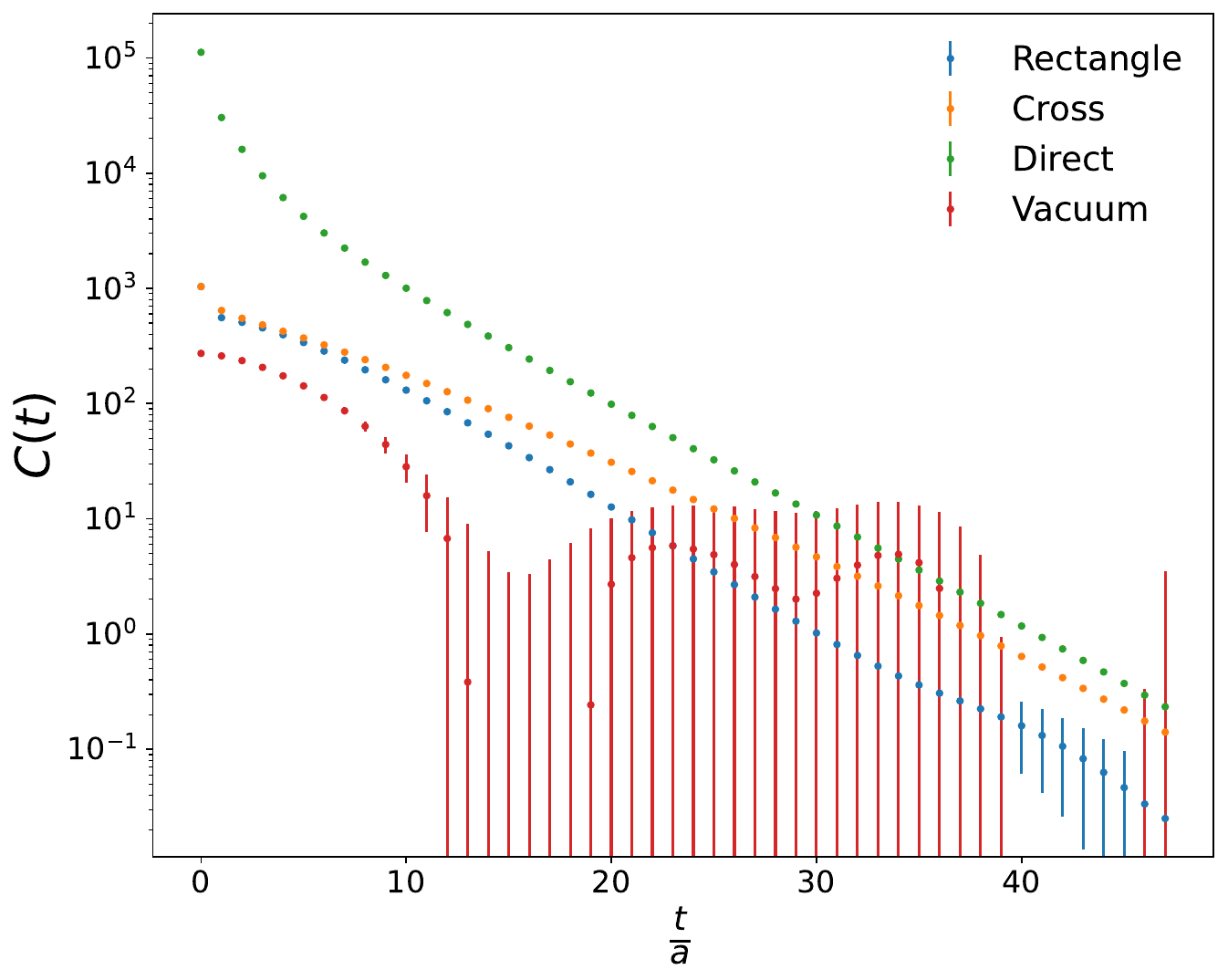}
  \caption{Different contributions arising from the Wick contractions for the two-pion correlator \eq{e:twopioncorr}. The Wick contractions are displayed in \fig{f:2piwick}. \label{f:2picorr}}
\end{figure}
\begin{figure}[t]
  \centering
  \includegraphics[width=0.15\textwidth]{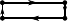}
  \hspace{0.05\textwidth}
  \includegraphics[width=0.15\textwidth]{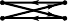}
  \hspace{0.05\textwidth}
  \includegraphics[width=0.15\textwidth]{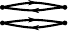}
  \hspace{0.05\textwidth}
  \includegraphics[width=0.15\textwidth]{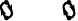}
  \caption{Wick contractions for the correlator of two pions in a singlet state. From left to right: Rectangle, Cross, Direct, Vacuum. \label{f:2piwick}}
\end{figure}
As an example of the quality of our data we show in \fig{f:2picorr} the terms that form the correlator between two-pion states
\be
\ev{ \Op_{2\pi}(t) \bar{\Op}_{2\pi}(0) } = 
- \frac{32}{3}\mathcal{R} + \frac{2}{3}\mathcal{C}  + 2 \mathcal{D} + 8\mathcal{V} \,.\label{e:twopioncorr}
\ee
Here $\mathcal{R}$ stands for Rectangle, $\mathcal{C}$ for Cross, $\mathcal{D}$ for Direct and $\mathcal{V}$ for Vacuum and denotes the four possible Wick contraction depicted in \fig{f:2piwick}. For now we computed the correlators using standard distillation without profiles. The Vacuum correlator is a disconnected quark-line contribution and has the largest statistical noise. We are developing methods based on multi-level sampling to ameliorate the signal-to-noise ratio in such cases \cite{Barca:2025fdq}.
\begin{figure}[t]
  \centering
  \includegraphics[width=0.49\textwidth]{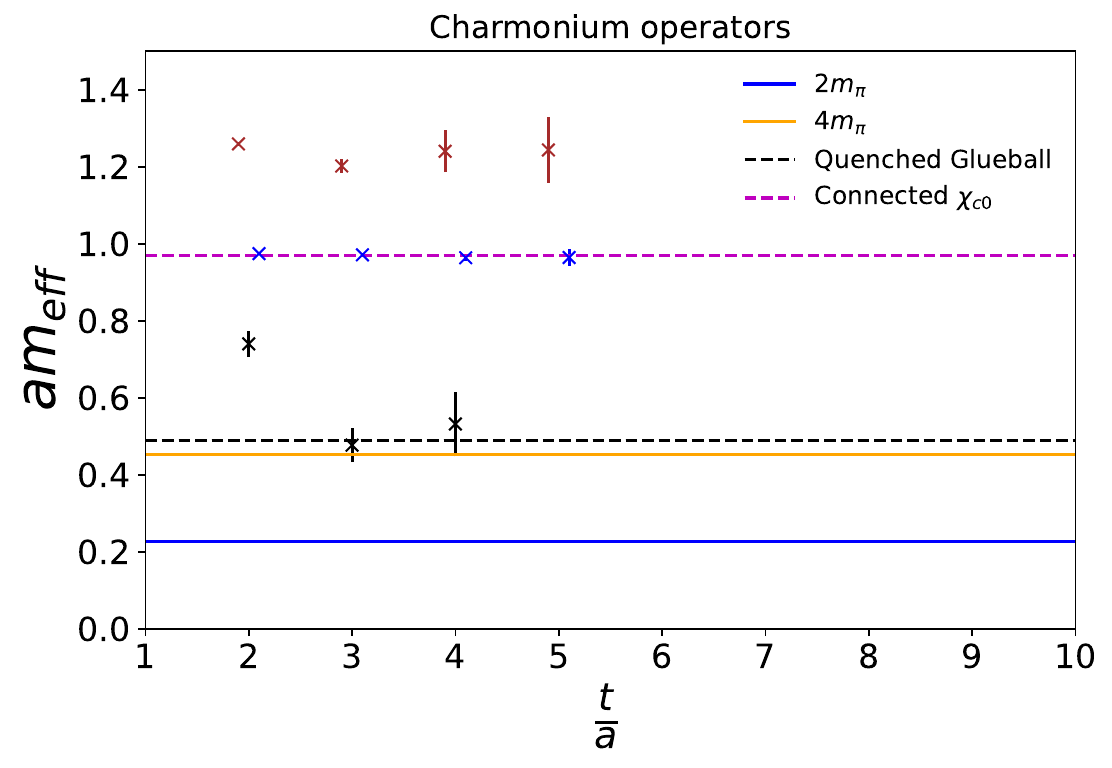}
  \includegraphics[width=0.49\textwidth]{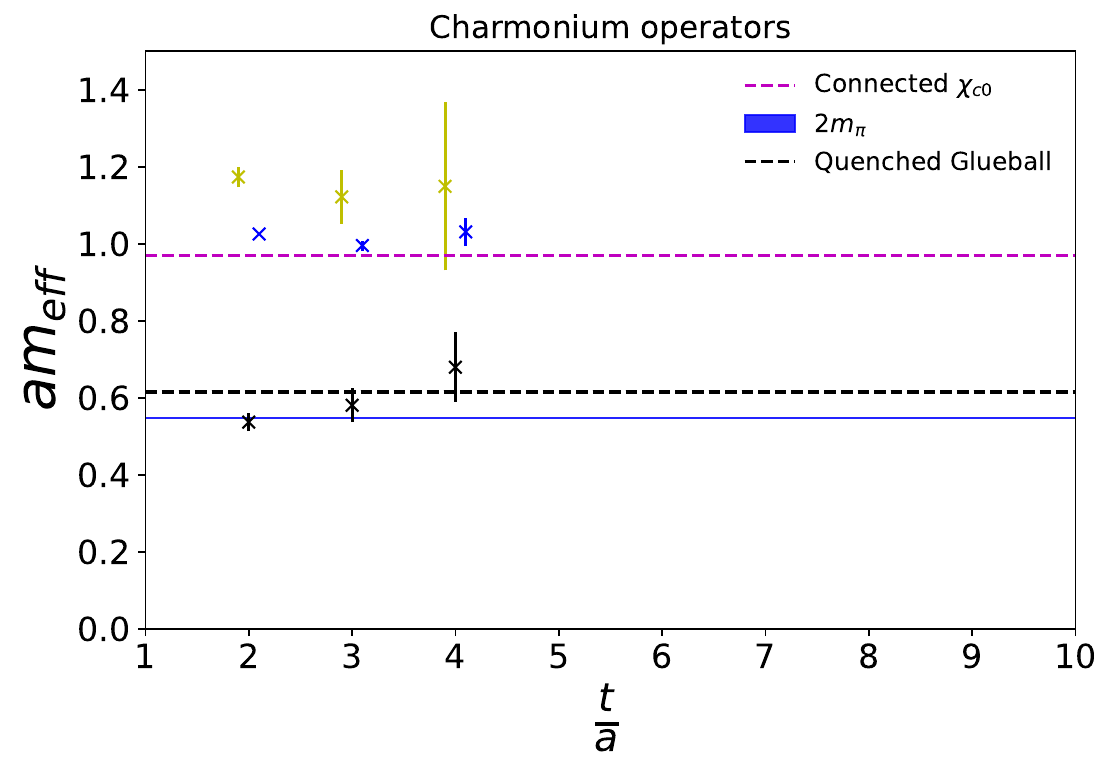}
  \caption{Effective masses from charmonium operators for ensemble A1 (left) and A1h (right). From \cite{Urrea-Nino:2025ztj}. \label{f:charmonium}}
\end{figure}

After pruning the correlation matrix \eq{e:corrmatrix} to 3 charmonium operators, 5 light meson operators and keeping one two-pion and one gluonic operator, we solve the GEVP, see \cite{Urrea-Nino:2025ztj}. We compute the effective masses for state $n$ by using the vectors $v_n$ computed at a fixed time to extract (approximations to) the eigenvalues $\lambda_n(t)=v_n^\dagger C(t) v_n / v_n^{\dagger} C(t_{\rm ref}) v_n$, cf. \eq{e:gevp} and from those the effective masses
\be
am_{{\rm eff},n}(t) = \ln\left(\frac{\lambda_n(t)}{\lambda_n(t+a)}\right)\,.
\ee
\fig{f:charmonium} shows the effective masses we obtain from a GEVP with the block of the correlation matrix corresponding to the three charmonium operators only. One the left are the results for the ensemble A1 and on the right for the heavier pion ensemble A1h. In the plots we also show lines corresponding to the quenched glueball mass (black dashed line), the threshold of two non-interacting pions at rest (blue line) and the lowest charmonium state $\chi_{c0}$ neglecting the effects of disconnected contributions (magenta dashed line). For the ensemble at lighter pion mass we also display the four-pion threshold (yellow line). We observe that the lowest resolved state is below the charmonium level (expected close to the connected-only level) and arises because we have included the disconnected contributions. The charmonium state appears as the first excited state.
\begin{figure}[t]
  \centering
  \includegraphics[width=0.49\textwidth]{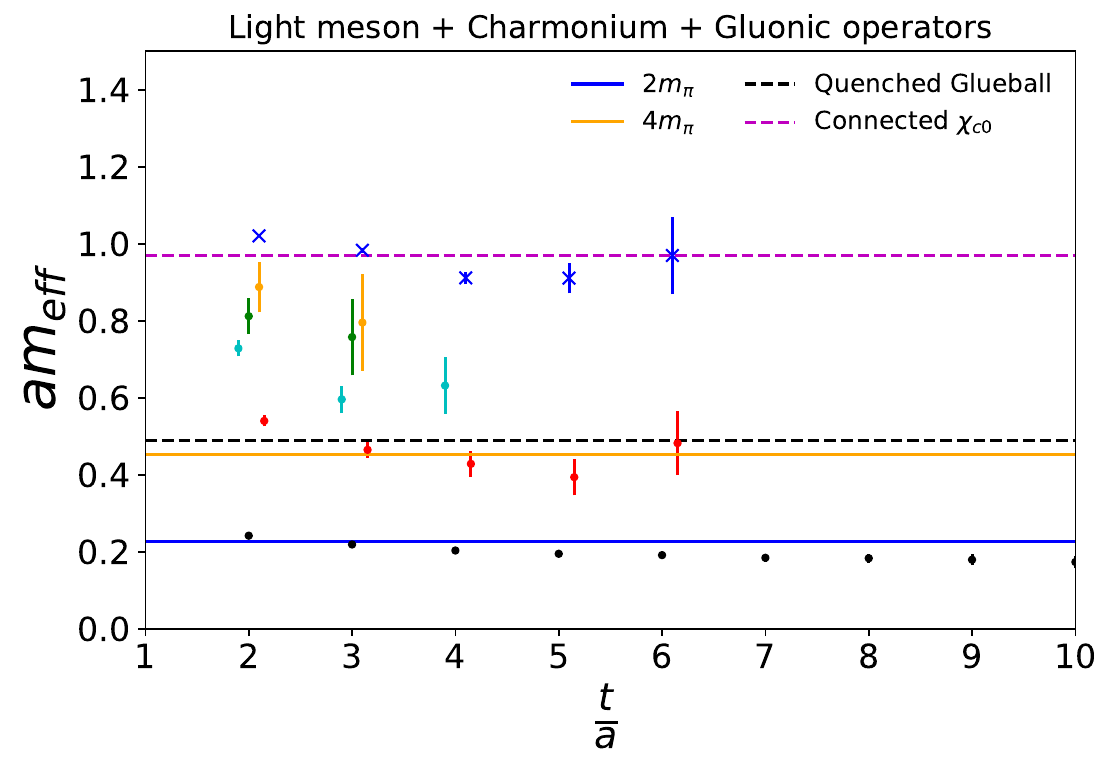}
  \includegraphics[width=0.49\textwidth]{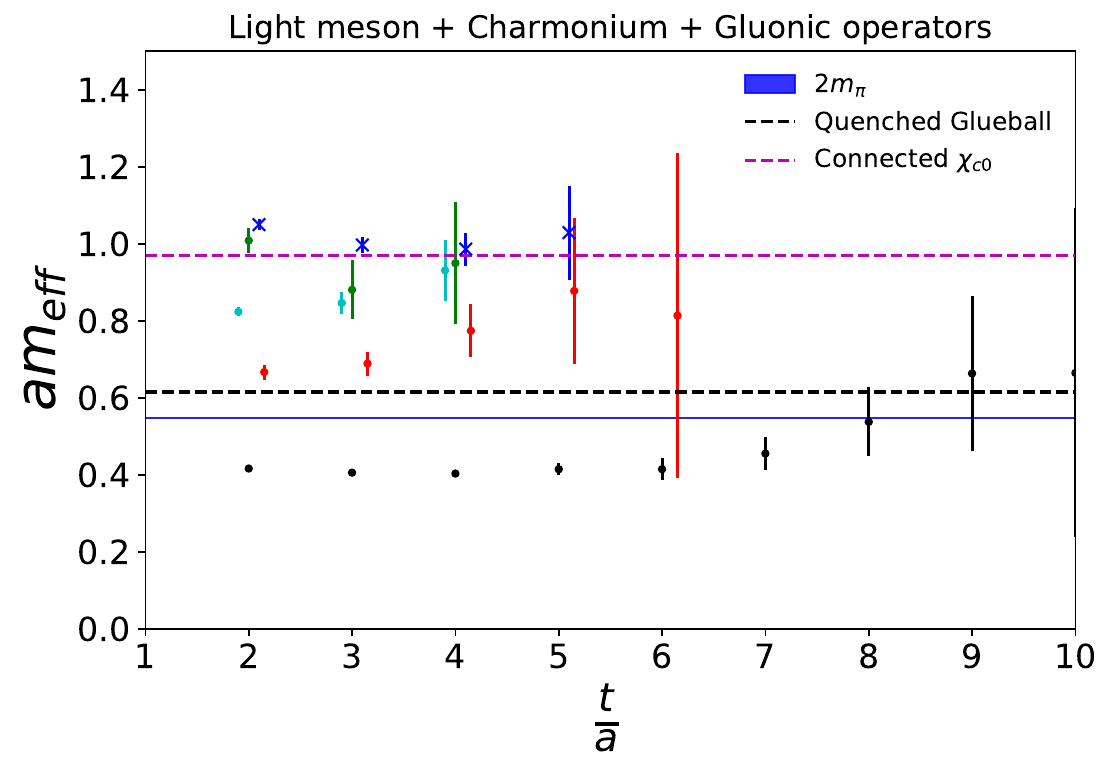}
  \caption{Effective masses from charmonium, light meson and glueball operators for ensemble A1 (left) and A1h (right). From \cite{Urrea-Nino:2025ztj}. \label{f:charmlightglue}}
\end{figure}

\fig{f:charmlightglue} shows the effective masses we obtain from a GEVP with the matrix correlation including 5 light one-particle meson, 3 charmonium and one glueball operators. As it was noted in \cite{Urrea-Nino:2025ztj} the addition of the glueball operator to the basis of light mesons and charmonium operators does not lead to the appearance of a new state in the low-lying part of the spectrum displayed in the figure. This was also observed in \cite{Brett:2019tzr}. The ground state (black points) corresponds probably to the $\sigma$ meson, as can be inferred by looking at the overlaps, see \fig{f:fullmixing} below. The first excited state (red points) appears close to the energy that corresponds to the quenched glueball on both ensembles.
\begin{figure}[t]
  \centering
  \includegraphics[width=0.49\textwidth]{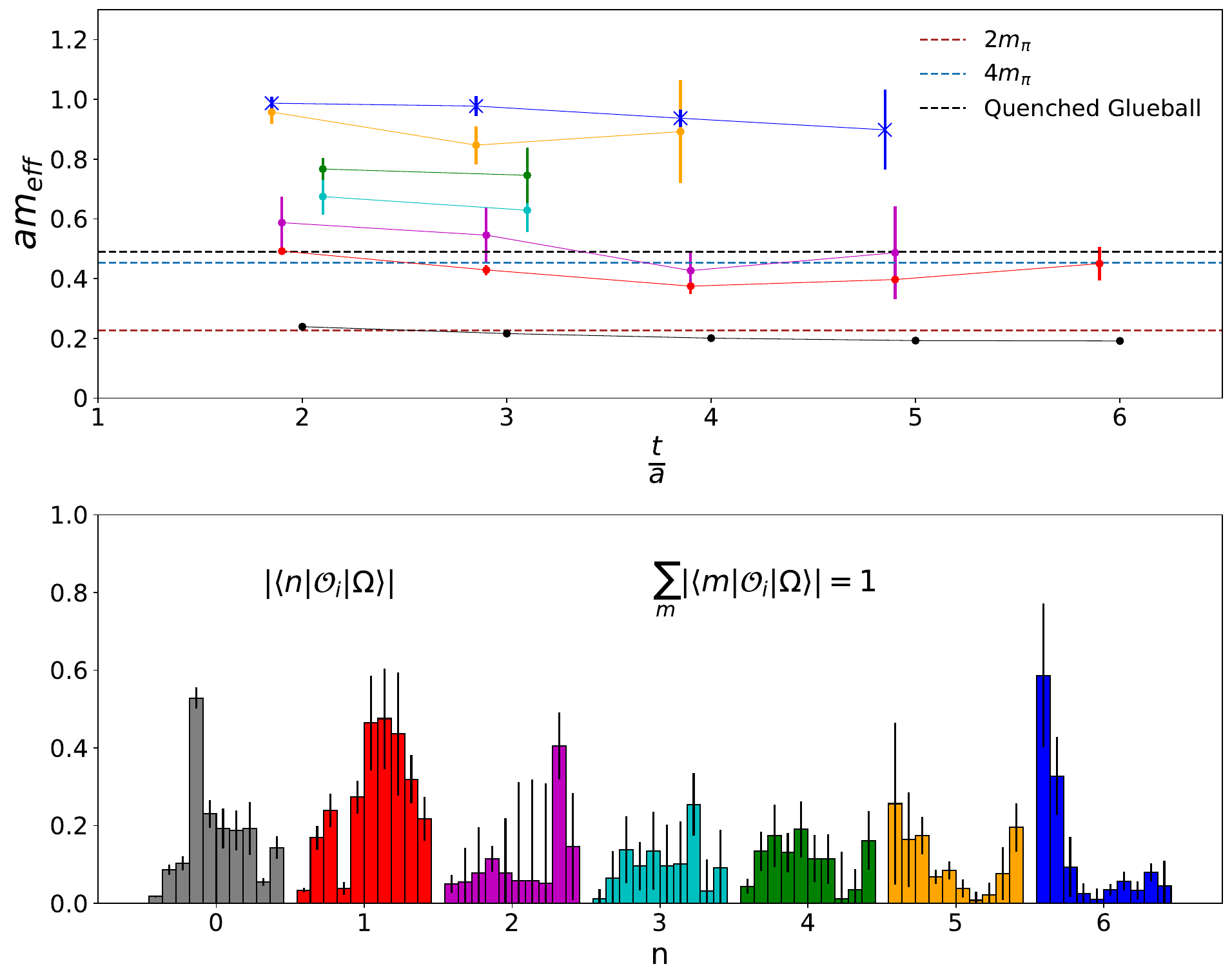}
  \includegraphics[width=0.49\textwidth]{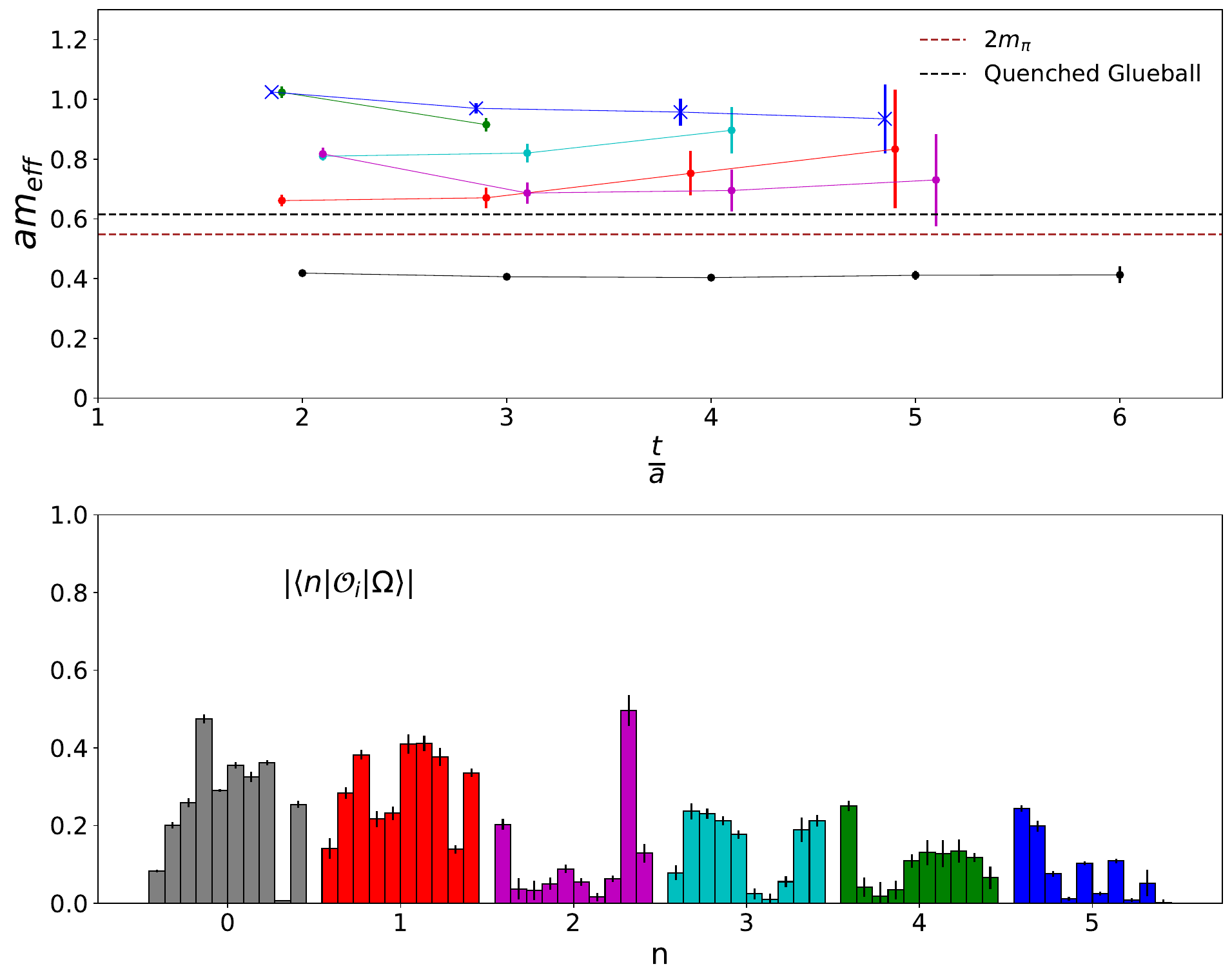}
  \caption{(Top row) Effective masses from charmonium, light meson, glueball and two-pion (both at rest) operators for ensemble A1 (left) and A1h (right). (Bottom row) For each eigenstate $n$ we show the overlaps with the operators $\Op_i$ in the basis. The normalization is given by \eq{e:ovnorm}. For each state $n$ the bars correspond in order to 3 charmonium operators, 5 light meson operators, one 2-pion operator and one gluonic operator. From \cite{Urrea-Nino:2025ztj}. \label{f:fullmixing}}
\end{figure}

The effective masses obtained with the full correlation matrix are shown in the top row of plots in \fig{f:fullmixing}. The lines are drawn to guide the eye. The addition of the two-pion operator to the basis leads to the appearance of a new state (magenta points) compared to what we see in \fig{f:charmlightglue}. This state appears for both pion masses close to the quenched glueball energy (black dashed line) and to the first excited state (red points). In the bottom row of plots in \fig{f:fullmixing} we show histograms of the overlaps \eq{e:overlap}. We normalize the overlaps such that their absolute values sum up to one for a given operator $\Op_i$
\be
\sum_n|\langle n|\Op_i|\Omega\rangle|=1 \,.\label{e:ovnorm}
\ee
The sum extends over the resolved states $n$. The state $n=1$ (red) has significant overlaps with all types of operators. The state $n=2$ (magenta) has predominantly overlap with the state created by the two-pion operator.

\section{Conclusions}

We computed the mixing of light meson, charmonum, glueball and two-pion at zero momentum states in the flavor singlet scalar channel through lattice simulations of $N_{\text f}=3+1$ QCD at two values of the pion mass. We observe that the inclusion of disconnected diagrams in charmonium correlators leads to a state in the region of light mesons, see \fig{f:charmonium}. The addition of the two-pion operator to the basis leads to the appearance of a new (second excited) state with respect to the spectrum obtained with one-particle operators only. This new state is close to the first excited state and they both appear near the quenched $0^{++}$ glueball energy, see \fig{f:fullmixing}. The inclusion of glueball operators in the basis does not lead to a new state that was not already present with light meson and charmonium operators. The statistical noise from glueball and disconnected correlators is a major problem which can be addressed by multi-level sampling methods \cite{Ce:2016ajy}. Moreover we will extend the basis to include two-pion operators at non-zero momentum in order to have a more complete spectrum of states.

\acknowledgments I thank my collaborators T. Korzec and J.A. Urrea-Niño for their help with this presentation.
The authors gratefully acknowledge the Gauss Centre for Supercomputing
e.V. (www.gauss-centre.eu) for funding this project by providing computing time on the GCS
Supercomputer SuperMUC-NG at Leibniz Supercomputing Centre (www.lrz.de) under GCS/LS
project ID pn29se as well as computing time and storage on the GCS Supercomputer JUWELS
at Jülich Supercomputing Centre (JSC) under GCS/NIC project ID HWU35. The authors also
gratefully acknowledge the scientific support and HPC resources provided by the Erlangen National
High Performance Computing Center (NHR@FAU) of the Friedrich-Alexander-Universität
Erlangen-Nürnberg (FAU) under the NHR project k103bf. M. P. was supported by the European
Union’s Horizon 2020 research and innovation programme under grant agreement 824093
(STRONG-2020). R.H. was supported by the programme "Netzwerke 2021", an initiative of the
Ministry of Culture and Science of the State of NorthrhineWestphalia, in the NRW-FAIR network,
funding code NW21-024-A. J. F. acknowledges financial support by the Next Generation Triggers project
(https://nextgentriggers.web.cern.ch). J. A. Urrea-Niño is supported by the German Research Foundation (DFG) research unit FOR5269 "Future methods for studying confined gluons in QCD" and also acknowledges financial support by the Inno4scale project, which received funding from the European High-Performance Computing Joint Undertaking (JU) under Grant Agreement No. 101118139.

%
\bibliographystyle{JHEP}
\bibliography{cairns}

\end{document}